\documentclass[12pt]{iopart}
\usepackage{color}
\usepackage{graphicx}
\usepackage{amsfonts}
\usepackage{amssymb}
\usepackage{cite}
\usepackage{iopams}
\begin{document}
\title{Storage of up-converted telecom photons in a doped crystal}
\author{Nicolas Maring$^1$, Kutlu Kutluer$^1$, Joachim Cohen$^{1,2}$, Matteo Cristiani$^1$, Margherita Mazzera$^1$, Patrick M. Ledingham$^1$, Hugues de Riedmatten$^{1,3}$}
\address{{$^1$ ICFO-Institut de Ciencies Fotoniques, Mediterranean Technology Park, 08860 Castelldefels (Barcelona), Spain.\\
$^2$ Current address: INRIA Paris-Rocquencourt, Domaine de Voluceau, B.P. 105, 78153 Le Chesnay Cedex, France.\\
$^3 $ ICREA-Instituci\' o Catalana de Recerca i Estudis Avan\c cats, 08015 Barcelona, Spain.\\}}
\ead{patrick.ledingham@icfo.es}
\begin{abstract} 
We report on an experiment that demonstrates the frequency up-conversion of telecommunication wavelength single-photon-level pulses to be resonant with a $\mathrm{Pr}^{3+}$:$\mathrm{Y}_2\mathrm{Si}\mathrm{O}_5$ crystal. We convert the telecom photons at $1570\,\mathrm{nm}$ to $606\,\mathrm{nm}$ using a  periodically-poled potassium titanyl phosphate nonlinear waveguide. The maximum device efficiency (which includes all optical loss) is inferred to be $\eta_{\mathrm{dev}}^{\mathrm{max}} = 22 \pm 1\,\%$ (internal efficiency $\eta_{\mathrm{int}} = 75\pm8\,\%$) with a signal to noise ratio exceeding 1 for single-photon-level pulses with durations of up to 560$\,$ns. The converted light is then stored in the crystal using the atomic frequency comb scheme with storage and retrieval efficiencies exceeding $\eta_{\mathrm{AFC}} = 20\,\%$ for predetermined storage times of up to $5\,\mu\mathrm{s}$. The retrieved light is time delayed from the noisy conversion process allowing us to measure a signal to noise ratio exceeding 100 with telecom single-photon-level inputs. These results represent the first demonstration of single-photon-level optical storage interfaced with frequency up-conversion.
\end{abstract}


\section{Introduction}

Quantum information technology relies on quantum light-matter interfaces, in particular, quantum memories for light \cite{Bussieres2013}. Such a device would allow for the synchronization of probabilistic events, for example generation of photon pairs from independent sources \cite{Nunn2013}. This is an important resource in the context of linear-optical quantum computation \cite{Knill2001, Kok2007} and long distance quantum communication with quantum repeaters \cite{Briegel1998,Duan2001}. This architecture promises to overcome the unavoidable problem of the limited distance due to loss of fiber-based quantum communication schemes  \cite{Sangouard2011}. 

Quantum photonic memories utilize atomic systems, as light can be absorbed on the optical transitions and stored as atomic excitations for durations limited by the coherence time. Moreover, if this coherence can be transferred to the ground state as a spin-wave, even longer storage times are possible. Many atomic systems have been investigated as quantum memories for light, including single atoms \cite{Specht2011}, warm atomic vapours \cite{Hosseini2011, Reim2011}, laser cooled atomic ensembles \cite{Chaneliere2005, Chou2005, Radnaev2010, Bao2012}, room temperature bulk diamond \cite{England2013}, nitrogen-vacancy centers in diamond \cite{Grezes2014}, hydrogen molecules \cite{Bustard2013}, metastable helium \cite{Maynard2014} and rare earth ion doped solids at cryogenic temperatures \cite{Riedmatten2008, Hedges2010, Lauritzen2010, Afzelius2010, Clausen2011, Saglamyurek2011, Lauritzen2011, Gundogan2012, Bonarota2013, Sabooni2013, Heinze2013, Dajczgewand2014, Bussieres2014, Sinclair2014, Saglamyurek2014b}.

Of great interest to long distance quantum communication is the implementation of a quantum memory for telecommunication light (1550$\,$nm). It is well known that telecom C-band optical fibers have the least loss (0.2$\,$dB/km), so a telecom optical memory would integrate directly with fiber-optical networks and  quantum repeater architectures. The obvious strategy for a telecom memory is the use of an atomic system which can directly absorb telecom light, for example erbium-doped solids which have a transition around 1530$\,$nm. Unfortunately, for this material (at a few Kelvin) the degree of spin polarization is low and thus limits the overall efficiency for memories based on spectral tailoring of the inhomogeneously broadened line. Two memory protocols were investigated in $\mathrm{Er}^{3+}$:$\mathrm{Y}_2\mathrm{Si}\mathrm{O}_5$ with light below the single-photon-level \cite{Lauritzen2010, Lauritzen2011}, the atomic frequency comb (AFC) \cite{Afzelius2009} and controlled reversible inhomogenous broadening (CRIB) \cite{Moiseev2001, Nilsson2005} protocols, with sub-percent total efficiencies at sub-microsecond storage times. Further to these experiments, non-classical telecom light storage in an erbium-doped optical fiber using the AFC protocol was recently demonstrated \cite{Saglamyurek2014b}, with efficiencies of about $1\,\%$ at a storage time of $5\,$ns.

Other protocols that do not require spectral tailoring have been proposed recently \cite{Damon2011, McAuslan2011}. In particular, a memory efficiency of $40\,\%$ was observed with strong input pulses and $\mathrm{Er}^{3+}$:$\mathrm{Y}_2\mathrm{Si}\mathrm{O}_5$ \cite{Dajczgewand2014} using the revival of silenced echo (ROSE) protocol. However, since these protocols are based on strong rephasing pulses at the same frequency as the retrieved echo, the ability to use them as a quantum memory remains technically challenging. In a different system, the ROSE protocol was limited to input weak coherent states of 14 photons per pulse, due to the noise created by the strong control pulses \cite{Bonarota2013}.

An alternate option with a telecom transition are rubidium-based systems, where the use of non-linear processes, such as  four-wave-mixing (FWM), can convert the frequency of the input telecom photons to be resonant with the rubidium based memory and has recently been demonstrated in \cite{Ding2012} albeit with the use of strong input light and a very low efficiency.

Another strategy implemented in \cite{Bussieres2014} is to teleport the quantum state of a telecom photon into the memory using entanglement between a telecom photon and a collective atomic excitation \cite{Clausen2011, Saglamyurek2011}. This method benefits in that any memory can be used, provided one can herald a photon pair where one of the pair is resonant with the memory and the other at telecom frequency. Then, the quantum state of an additional telecom photon can be teleported into the memory by performing a Bell-state measurement between said photon and the generated telecom photon.

Yet another alternative is to use quantum frequency conversion via a non-linear process as an interface between a memory and telecom wavelengths. So far, this has only been demonstrated for emissive type memories based on the Duan-Lukin-Cirac-Zoller (DLCZ) protocol \cite{Duan2001}. The signal Stokes photon has been down- and then up-converted to and from a telecommunication wavelength via FWM in an additional cold rubidium ensemble  \cite{Radnaev2010} and the heralded anti-Stokes photon has been frequency down-converted to telecom light using a nonlinear waveguide \cite{Albrecht2014}. For both cases, the non-classicality between Stokes and anti-Stokes was preserved.

Here, we report on the first demonstration of interfacing single-photon-level telecom pulses to an absorptive memory using frequency conversion. The material we use as the memory is $\mathrm{Pr}^{3+}$:$\mathrm{Y}_2\mathrm{Si}\mathrm{O}_5$ with the optical transition at $606\,$nm, a particularly good candidate for a quantum optical memory. Included in the recent experiments with this system are the $56\,\%$ storage of light using the AFC protocol \cite{Sabooni2013}, $69\,\%$ quantum storage of weak coherent states using the CRIB protocol \cite{Hedges2010}, light storage for up to one minute using electromagnetically induced transparency \cite{Heinze2013}, quantum storage of polarization qubits \cite{Gundogan2012}, storage of quantum light \cite{Rielander2014} and spin-wave storage \cite{Afzelius2010, Gundogan2013}. To interface this material with the telecom C-band, we use frequency up-conversion in a nonlinear waveguide to convert weak light at 1570$\,$nm to $606\,$nm. Such a technique has been employed as an efficient way to detect broadband single-photon-level telecom light \cite{Vandevender2004, Roussev2004, Albota2004, Thew2006, Pelc2011, Pelc2012} and conversion of nonclassical light \cite{Tanzilli2005, Rakher2010, Ates2012, Vollmer2014}. However, in order to match our optical memory bandwidth, long photons $(>\,100\,\mathrm{ns})$ should be converted, which increases the requirements for noise suppression. For the maximum pump power available in our experiment, the device efficiency of the converter (inclusive of all optical loss) is measured to be $ 21\pm1\,\%$. Using the AFC protocol we can store the up-converted photon for times up to $\tau_{\mathrm{AFC}} = 10\,\mu\mathrm{s}$ with a maximum overall efficiency (conversion efficiency, memory efficiency and all optical loss) of $\eta_{\mathrm{tot}} = 1.9\pm0.2\,\%$ at $\tau_{\mathrm{AFC}} = 1.6\,\mu\mathrm{s}$ storage time, currently outperforming previous realizations of absorptive telecom optical memories at the single-photon-level.

\section{Experimental Set Up}

\begin{figure}[h]
\centering\includegraphics[width=1.05\textwidth]{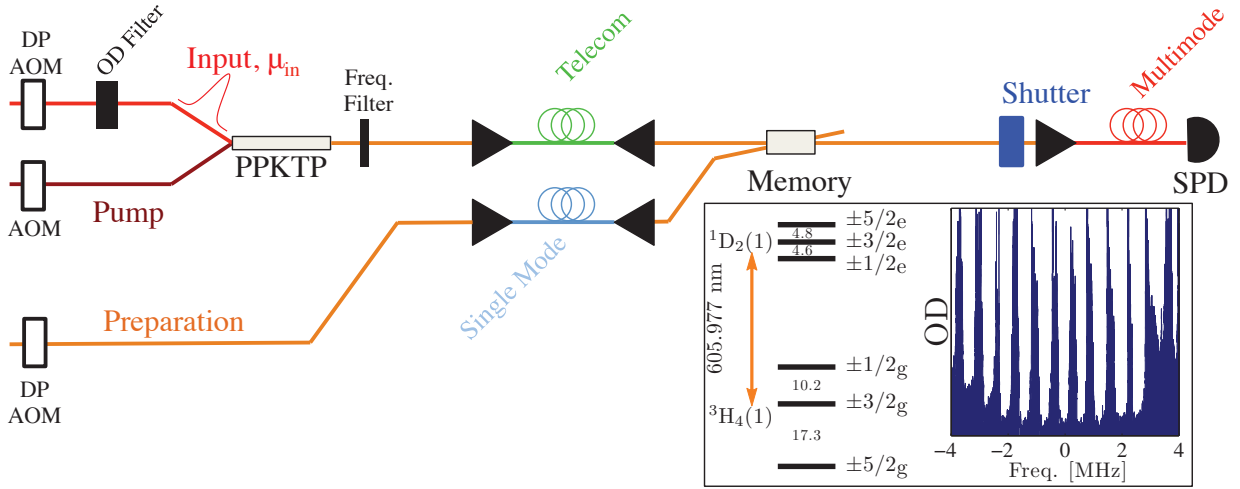}
\caption{Experimental set-up. Pulses of $1570\,$nm light are created with a double pass acousto-optic modulator (DP AOM) and are then reduced to the single-photon-level using a neutral density filter (OD Filter). Both signal and pump at $987\,$nm are coupled into the periodically-poled potassium titanyl phosphate (PPKTP) nonlinear waveguide producing the $606\,$nm output. The output mode of the FC is filtered with first a diffraction grating and then an etalon with finesse 6, linewidth $10\,$GHz. The mode is then coupled to the memory based on the AFC protocol (see text) via a telecom fiber (slightly multimode for $606\,$nm). The light used to create the AFC passes a DP AOM before being in and out coupled of a single mode optical fiber, giving a total of $2\,$mW of power before the crystal. The AFC output mode is coupled to a single photon detector via a multimode fiber. A shutter is used to protect the SPD during the preparation period. Inset: Left: The level scheme for $\mathrm{Pr}^{3+}$:$\mathrm{Y}_2\mathrm{Si}\mathrm{O}_5$  with the hyperfine splittings given in MHz. Right: an example of the AFC for a storage time of $\tau_\mathrm{AFC} = 1.6\,\mu\mathrm{s}$. Note that the plot is indicative only as the measurement of the optical depth is limited by the dynamic range of the detection system used. This AFC is resonant with the $\pm 1/2_\mathrm{g} - \pm 3/2_\mathrm{e}$ transition.} \label{fig:1}
\end{figure}

Figure \ref{fig:1} shows a simplified experimental set-up. There are two main parts to the device: the frequency converter (FC) and the AFC solid-state optical memory, of which we first describe the former. 

\subsection{Frequency Converter\label{subsec:converter}}

Our FC is based on a frequency up-conversion process in a 2.6$\,$cm long periodically-poled potassium titanyl phosphate (PPKTP) nonlinear waveguide (AdvR Inc.). With pump photons at $987\,$nm and signal photons at $1570\,$nm coupled to the waveguide, a nonlinear process frequency up-converts the telecom photons to 606$\,$nm when both energy and momentum are conserved \cite{Boyd1992}. The waveguide has a quasi-phase matched bandwidth over which efficient frequency conversion occurs, the bandwidth depending on the poling period \cite{Boyd1992}. 

For the pump, we use a Toptica external cavity diode laser (ECDL) with a tapered amplifier (TA)  providing about 1.2$\,$W of $987\,$nm, which is coupled to a single mode fiber resulting in a maximum of 600$\,$mW at the input of the FC. We note that some of this light is also used to generate the 606$\,$nm beam for preparation of the memory (see subsection \ref{subsec:memory}). For the signal, we use again a Toptica ECDL but with a Keopsys erbium-doped fiber amplifier (EDFA) to have around $1\,$W of $1570\,$nm. A 95:5 fiber beam splitter is used, the weaker port used for the FC, the stronger port used to create light for the memory preparation (see subsection \ref{subsec:memory}).

To couple to the nonlinear waveguide, we use a lens (Thorlabs, f~=~$11\,$mm, NA~=~$0.25$) mounted on a 3-axis translation stage. The transmission in the waveguide (including waveguide coupling and losses) of the pump and signal modes are measured to be $36\,\%$ and $55\,\%$, respectively, both cases being limited by mode-matching. Both in-coupled optical modes pass through acousto-optic modulators (AOM - AA Optoelectronic). For the signal mode, there is a double-pass configuration which ensures the frequency of the up-converted photons exactly matches that of the memory (see subsection \ref{subsec:memory}). Also in the signal mode are OD filters to reduce the light to the single-photon-level. The AOM in the pump mode is operated in single pass and acts as a gate, the zeroth order mode being coupled to the FC such that when the AOM is activated, about $75\,\%$ of the pump power is steered away from the FC. Moreover, when the AOM is activated the mode shape of the zeroth order is distorted, further reducing the coupled power to the waveguide. With this gating system, we are able to reduce the up-converted signal by a factor exceeding 10. This gating becomes useful for detection of the light retrieved from the crystal.

The up-converted mode is subject to frequency filtering. We use a diffraction grating (Thorlabs, GR13-1205)  to remove from the output mode the pump light, the signal light that is not converted and the up-converted pump light ($493.5\,$nm). A more problematic source of noise is pump induced Raman scattering and spurious spontaneous parametric down-conversion \cite{Pelc2011}. This noise can include the input frequency and will thus be up-converted as a broadband noise about the target frequency, the bandwidth given by the quasi-phase matching condition. To reduce this effect, we employ an etalon (Light Machinery) with finesse $6$ and linewidth $10\,$GHz. The total transmission from the output of the waveguide to after the last frequency filtering component is $71\,\%$.

Finally, the output mode is coupled to a telecommunication wavelength fiber which is slightly multimode for 606$\,$nm but allows for an efficient coupling of $75\,\%$. The combination of this fiber with the grating gives an estimated filtering bandwidth of $80\,$GHz. The light is then either coupled to a single photon detector for direct analysis of the frequency converter, or coupled to the crystal for storage. The total transmission from the output of the waveguide to after the fiber is then $53\,\%$.

\subsection{Storage Device\label{subsec:memory}}

Our device is based on the atomic frequency comb (AFC) protocol \cite{Afzelius2009}. This protocol is well suited for inhomogeneously broadened and spectral holeburning media. An incoming photon is collectively absorbed by the medium creating an atomic coherence which dephases due to the inhomogeneity. However, if the absorption line is tailored into a periodic structure with a well defined frequency spacing of $\Delta$ (referred to as a comb), the atomic coherence will rephase at a time $\tau_\mathrm{AFC} = 2\pi/\Delta$ resulting in photon-echo like emission. For a given storage time and optical depth, there is an optimal finesse, $F$, that maximizes the efficiency, where the finesse is defined as the ratio of the comb teeth spacing $\Delta$ to the comb tooth width $\gamma$. Additionally, there is a lineshape of the teeth of the comb that maximizes the efficiency \cite{Chaneliere2010}. 

We note that the storage time here is predetermined and therefore the light can not be recalled on-demand. However, the full AFC scheme is directly applicable in our system \cite{Afzelius2010, Gundogan2013}, where the optical excitation can be transferred to the ground state to be stored as a long-lived spin-wave and then recalled on-demand. Such on-demand memories become useful as synchronizing devices, e.g. in the context of quantum repeater technology \cite{Briegel1998}. In addition, note that the AFC scheme alone can be useful even with a fixed storage time, e.g. if combined with temporal and spectral multiplexing with feed-forward control and read-out in the spectral domain, as described and demonstrated in \cite{Sinclair2014}. Our system is compatible with this technique.

As the memory medium, we use a $5\,$mm long $\mathrm{Pr}^{3+}$:$\mathrm{Y}_2\mathrm{Si}\mathrm{O}_5$ crystal with $0.05\,\%$ doping leading to an absorption coefficient of $23\,$cm$^{-1}$ with light polarized along the D$_2$ crystal axis. For light polarized along the D$_1$ axis, the absorption coefficient is reduced greatly \cite{Sun2005}. The transition of interest is at $606\,$nm which connects the first sublevels of the $^3$H$_4$ ground and the $^1$D$_2$ excited manifolds. To cool this crystal to cryogenic temperatures we use a Montana Instruments Cryostation mounted on a different table to that of the laser source and conversion set-up. The $606\,$nm laser source used for preparation is based on sum-frequency-generation of 1570$\,$nm and $987\,$nm in a  second PPKTP waveguide. The light is derived from the same lasers described in subsection \ref{subsec:converter}. The frequency of the $987\,$nm light is locked via Pound-Drever-Hall to a cavity using 606$\,$nm light generated in a different waveguide set-up \cite{Gundogan2013}.  A single-mode fiber brings about 2$\,$mW of light to the Cryostation for the preparation mode.

We use a double pass AOM to first create a $12\,$MHz wide transparency window (referred to as `pit' ) via spectral holeburning the atoms to other hyperfine levels of $\mathrm{Pr}^{3+}$:$\mathrm{Y}_2\mathrm{Si}\mathrm{O}_5$. The population lifetime of the ground state is around $\sim 100\,$s \cite{Ohlsson2003a} with an optical excited state lifetime of $164\,\mu\mathrm{s}$ \cite{Equall1995}, making spin polarization and thus spectral holeburning simple. We burn ions back into the pit by sweeping the laser $4\,$MHz centered at $+30\,$MHz with respect to the pit centre. This populates the  $\pm\,1/2_\mathrm{g}$ and $\pm\,3/2_\mathrm{g}$ ground states within the pit (see inset of fig. \ref{fig:1} for level diagram), and by then emptying again the  $\pm\,3/2_\mathrm{g}$  ground state via optical pumping, we are left with a $4\,$MHz wide single class of ions within the $\pm\,1/2_\mathrm{g}$ ground state \cite{Nilsson2004, Guillot-Noel2009}. Within this single class, spectral holes at regular $\Delta$ intervals are burnt resulting in a $4\,$MHz wide AFC resonant with the $\pm\,1/2_\mathrm{g} - \pm\,3/2_\mathrm{e}$ transition (see inset of fig. \ref{fig:1}). The burnt ions are then placed into the $\pm\,3/2_\mathrm{g}$ and $\pm\,5/2_\mathrm{g}$ ground states. The total preparation time for the memory is around $200\,$ms.

Finally, the preparation mode is sent at an angle of about $4\,^{\circ}$ with respect to the input mode, overlapping only on the memory. This allows to spatially filter the strong preparation light away from the single photon detection mode.

Important for the overall efficiency of the device is the loss the input mode experiences from being out-coupled from the fiber, passing through the memory and being fiber coupled again to the detector. To reduce the loss we use a multimode fiber for coupling to the detector. The total transmission from after the telecom fiber to after the multimode fiber is $66\,\%$.

\subsection{Detection}

Further to spatially filtering the preparation light from the detection, a mechanical shutter is used to physically block the single photon detector (SPD) from leaked scattered light of the preparation. The SPD used is from PicoQuant (Tau SPAD-20) with a measured $10\,$Hz dark count rate and a specified $60\,\%$ detection efficiency at $606\,$nm. When coupling light after the memory to the detector we use a multimode fiber. When testing only the FC, the telecom fiber is used to couple to the SPD. Finally, as an extra filtering stage, we use a bandpass filter of $10\,$nm width around $605\,$nm which is placed directly before the fiber coupler in all detection cases.

With a time to digital converter (TDC - Signadyne), we perform start-stop measurements. We send $1.2 \times 10^6$ pulses at a rate of around $50\,\mathrm{kHz}$. When the memory is in place, these pulses are sent as 200 lots of 6000, where the 6000 pulses are sent within $150\,$ms. For each lot of 6000 pulses, a new AFC is prepared.

For all results shown, we note that the dark counts of the detector are not subtracted.

\section{Frequency conversion of single-photon-level light\label{sec:FC}}

\begin{figure}[h]
\centering\includegraphics[width=0.85\textwidth]{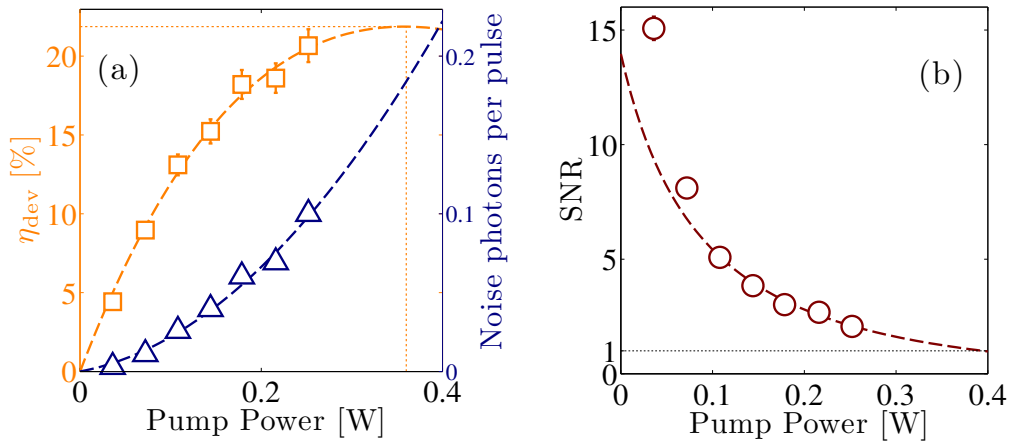}
\caption{\label{fig:2} Single-photon-level frequency conversion characteristics. (a) Device efficiency and noise photons per pulse at the detector versus pump power measured after the waveguide. The light orange squares (dark blue triangles) represent the signal (noise) with the dashed lines as fits described in the text. All data were taken with an input of $\mu_\mathrm{in} = 1$. The integration window size for the input was $400\,$ns. The dotted horizontal and vertical lines indicate the maximum device efficiency $(22\pm1\,\%)$ and the corresponding pump power $360\,$mW, respectively. (b) The signal to noise ratio as a function of pump power after the waveguide. The data are represented as circles with a dashed line fit as described in the text. Error-bars are plotted on all data and represent one standard deviation.}
\end{figure}

We first characterize the frequency conversion of single-photon-level $1570\,$nm light independent of the memory. An important value of the FC is the device efficiency $\eta_{\mathrm{dev}}$, which we define as the ratio of the number of $606\,$nm photons at the detector to the number of $1570\,$nm photons at the input lens of the device. We measure this as a function of pump power after the waveguide, $P_p$, in fig. \ref{fig:2}(a). For this data we use an input of $\mu_{\mathrm{in}} = 1$, which refers to the average number of photons per pulse. The pulse duration is $140\,$ns and we correct only for the SPD efficiency $(60\,\%)$. The data are fitted with the following equation
\begin{equation}
\eta_{\textrm{dev}} = \eta_{\textrm{dev}}^{\textrm{max}}\,\sin^2 \left( L\,\sqrt{P_p \, \eta_n} \right),
\end{equation}
where $L$ is the length of the waveguide and $\eta_n$ is the normalized efficiency. This equation can be obtained by treating the pump light classically and having vacuum at the converted photon frequency at the input of the device. The maximum measured device efficiency is $21 \pm 1 \,\%$  for $P_p = 252\,$mW. Extracting from the fit, the maximum device efficiency is inferred to be $\eta_{\textrm{dev}}^{\textrm{max}} = 22 \pm 1\,\%$ for $P_{\mathrm{max}} = 360\,$mW and a normalized efficiency of $\eta_n = 100 \pm 10\, \%/\mathrm{W} \, \mathrm{cm}^2$. Correcting for all known losses, e.g. fiber coupling ($75\,\%$) and propagation losses ($71\,\%$), we can infer the external efficiency, i.e. the efficiency directly outside of the waveguide, to be $\eta_{\mathrm{ext}} = 41\pm 4\,\%$. Taking into account the coupling of the $1570\,$nm to the waveguide  ($55\,\%$), we infer the internal efficiency  to be $\eta_{\mathrm{int}} = 75\pm8\,\%$. That is, $75\,\%$ of what telecom light is coupled into the waveguide is converted.

In fig. \ref{fig:2}(a) we show also the dependence of the noise as a function of pump power after the waveguide for $\mu_{\mathrm{in}} = 1$. Here we fit with a quadratic function of the form $\alpha\,P_p^2 + \beta\,P_p + \mathrm{DC}$ where $\alpha/\beta$ are free parameters and DC is the dark count of the detector. This is the expected behaviour of this system, the noise being a result of two processes: (i) the pump inducing photons at the input frequency via weakly phase-matched down-conversion due to non-perfect poling \cite{Pelc2011a} which are then up-converted and (ii) spontaneous Raman scattering of the pump \cite{Pelc2011}. The noise due to down-conversion can be avoided simply if the pump frequency is lower than that of the input. However, to interface light near the telecom C-band ($1.5 - 1.6\, \mu$m) to a $\mathrm{Pr}^{3+}$:$\mathrm{Y}_2\mathrm{Si}\mathrm{O}_5$ based memory ($606\,$nm), the pump ($1016 - 975\,$nm) will always be more energetic than the input. Therefore, for our application, this noise via SPDC is unavoidable, unless a cascaded conversion scheme is used \cite{Pelc2012}.

Shown in fig. \ref{fig:2}(b) is the signal to noise ratio versus pump power after the waveguide, where SNR$\,=\frac{\mathrm{S} - \mathrm{N}}{\mathrm{N}}$ with S being the total number number of counts (including noise) and N being the noise counts. At $P_{\mathrm{max}}$, the signal to noise ratio still exceeds 1. For pump powers larger than this, we enter the regime where the up-converted photons are being down-converted again, leading to signal to noise ratios below 1.

\section{Storing up-converted single-photon-level telecom light}

\begin{figure}[h]
\centering\includegraphics[width=0.65\textwidth]{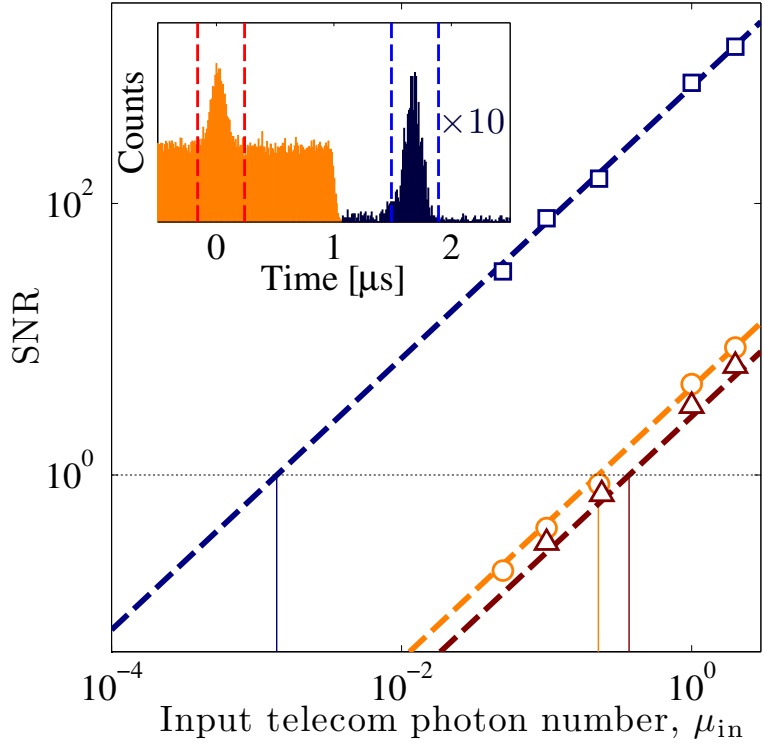}
\caption{\label{fig:3} Signal to noise ratio as a function of the input telecom photon number, $\mu_\mathrm{in}$ for a pump power of $144\,$mW after the waveguide. The integration window is 400$\,$ns. The red triangles show the results for the FC only, the orange circles show the case of the photons passing through a transparency window and finally the blue squares show the AFC echo. The dashed lines show a linear fit forced to go through zero. The black dotted line represents SNR$ = 1$, with the thin vertical lines showing the $\mu_1$ for each case. Error-bars (smaller than the symbols) represent one standard deviation. Inset: An example of an input (light, orange histogram) and an echo (dark, blue histogram) for $\mu_\mathrm{in} = 0.1$ photons per pulse and $\tau = 1.6\,\mu\mathrm{s}$. Note the echo histogram is multiplied by 10. The vertical dashed lines indicate the integration window for the signal. A comb is prepared 200 times, with 6000 pulses sent per comb. The bin size is $10.24\,$ns. }
\end{figure}

Having characterized the conversion, we now couple the light to the crystal, with fig. \ref{fig:3} showing the results. At the input of the FC, photon numbers ranging from $0.05$ to $2$ photons per $140\,$ns pulse were used and the signal to noise ratio was measured for 3 cases: the frequency converter alone, the photons passing through a $12\,$MHz transparency window created in the memory crystal and finally the stored and retrieved photons from the memory. For the second case, the light is polarized along the D$_1$ crystal axis, whereas for the third case the polarization is along the D$_2$ axis. The characteristic value we choose to compare for the 3 cases is the minimum input photon number required to achieve a signal to noise ratio of 1, the so-called $\mu_1$. The $\mu_1$ is extracted from a linear fit of the data, where the fit is forced to pass through zero. For all measurements in this section, the pump power is set to $144\,$mW leading to an FC device efficiency of $15\pm1 \,\%$.

For the case of the frequency converter alone, where the photons are measured after coupling to the telecom fiber, we measure ${\mu_1 = 0.37 \pm 0.02}$ photons. The integration window taken is $400\,$ns which includes the entire pulse. A $2.36\,\mu\mathrm{s}$ noise window is taken in the same trace, located $7.48\,\mu\mathrm{s}$ after the input pulse, with the counts normalized to $400\,$ns.

We then out-couple the  light and steer it toward a $12\,$MHz transparency window tailored in the inhomogenous absorption line of the $\mathrm{Pr}^{3+}$:$\mathrm{Y}_2\mathrm{Si}\mathrm{O}_5$ crystal and we measure ${\mu_1 = 0.23 \pm 0.004}$ photons. The signal and noise detection windows here are the same as the previous case. An improvement in the $\mu_1$ is seen, this is due to the absorption of the noise by inhomogenous linewidth of the sample. The etalon placed after the converter (see fig. \ref{fig:1}) has a linewidth of around $10\,$GHz resulting in noise with this bandwidth. With the noise polarized along the D$_1$ axis we measure a reduction of $33\,\%$ which is consistent with an absorbing line of width $6 - 12\,$GHz.

Finally, for the case of the stored light we measured ${\mu_1 = 1.38 \pm 0.03 \times 10^{-3}}$ photons, about a factor of $270$ improvement compared to the frequency converter only case. The memory efficiency is measured to be $\eta_\mathrm{AFC} = 19.8 \pm 0.1\,\%$. One key feature here is that the echo signal is delayed by a known time of $\tau_\mathrm{AFC} = 1.6\,\mu\mathrm{s}$, which allows us to disable the $987\,$nm pump of the frequency converter before the echo is re-emitted from the sample. This can be seen clearly in the inset of fig. \ref{fig:3} where at a time of about $1\,\mu\mathrm{s}$ the pump is disabled resulting in a dramatic drop of noise in the echo temporal window. The pump is disabled for a total time of $5\,\mu\mathrm{s}$, and to measure the noise we take a measurement with the input telecom photons physically blocked at the input of the converter and integrate over the entire $5\,\mu\mathrm{s}$ that the pump is off. The signal integration window remains the same as the previous cases.

Due to the narrowband nature of the memory (about $4\,$MHz relative to the $>5\,$ GHz noise), the memory itself acts as a filter of the noise storing only a fraction of it. This, in combination with the fact that the pump is gated, thus allows for a 2 orders of magnitude reduction in the $\mu_1$ for the echo. 

From the echo and the input photon number we can extract the total efficiency of our device (including frequency conversion, all optical loss and storage efficiency), which we measure to be $\eta_{\mathrm{tot}} = 1.55 \pm 0.02\,\%$. From the device efficiency seen in Sec. \ref{sec:FC} ($\sim15\,\%$ at 144$\,$mW pump power),  an additional $66\,\%$ transmission loss from memory input to the SPD and the AFC efficiency $(\sim 20\,\%)$, we have quantitative agreement between using strong and single-photon-level light.

\begin{figure}[h]
\centering\includegraphics[width=1\textwidth]{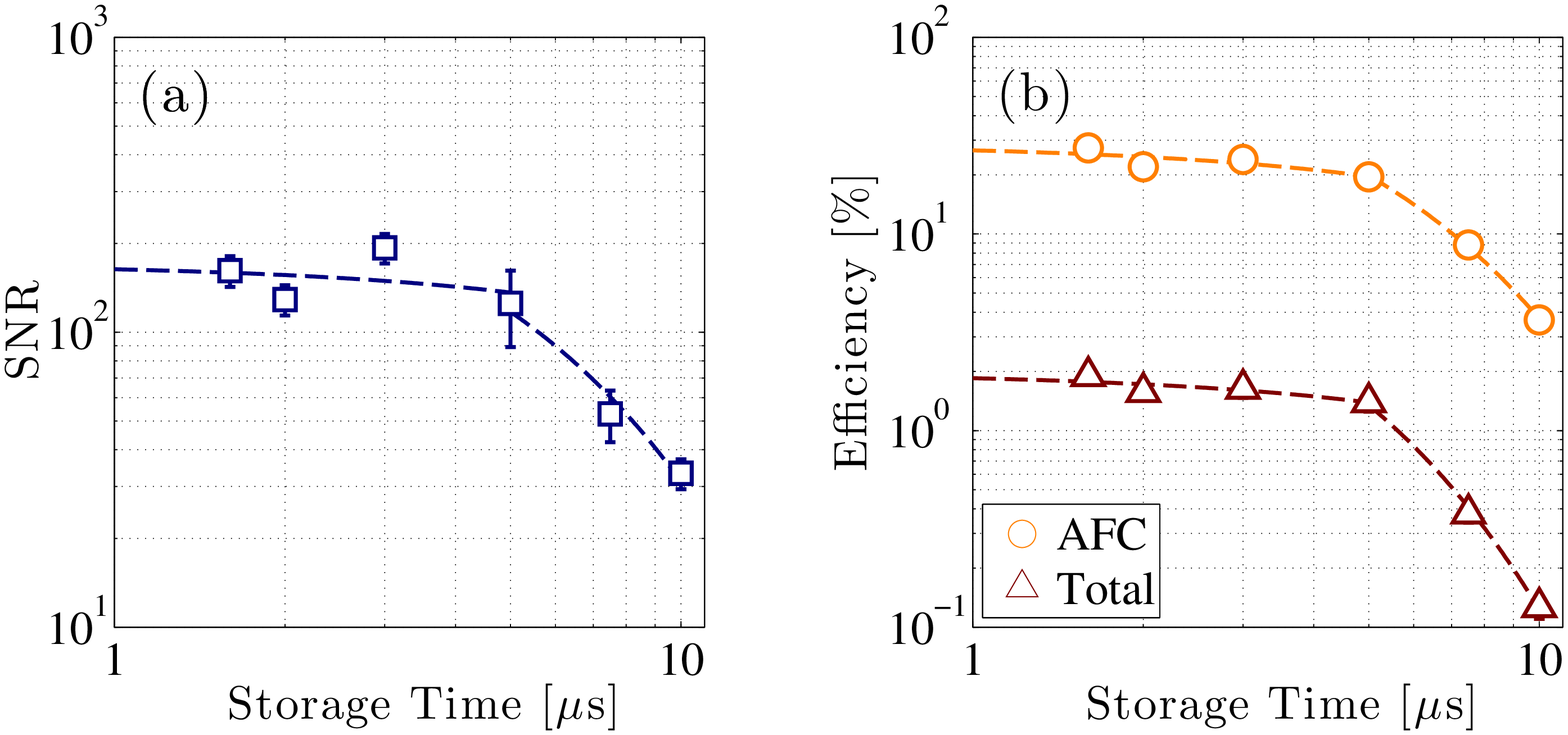}
\caption{\label{fig:4} (a) Signal to noise ratio and (b) echo/total efficiency vs storage time for $\mu_{\mathrm{in}} = 1$ telecom photons per 560$\,$ns pulse (excluding the data point at $10\,\mu$s where $\mu_\mathrm{in} = 0.4$ is used with the SNR normalized accordingly). The SNR and efficiencies stay relatively flat for storage times up to and including $5\,\mu$s. For these storage times, the total efficiency is always above $1\,\%$ and the SNR is always above 100 which would correspond to a $\mu_1$ lower than $10^{-2}$. The efficiency drops drastically to below $10\,\%$ at storage times longer than $7.5\,\mu$s, limited by the achievable comb tooth width $\gamma$. Dashed lines in both figures are used to guide the eye. Error-bars (often smaller than the symbol) represent one standard deviation.}
\end{figure}

We measure also the echo efficiency together with the signal to noise ratio as a function of storage time from $\tau_\mathrm{AFC} = 1.6\,\mu\mathrm{s}$ up to $\tau_\mathrm{AFC} = 10\,\mu\mathrm{s}$, shown in fig. \ref{fig:4}. For this data, we use pulses of an increased duration of 560$\,$ns while keeping the AFC bandwidth constant. We increase also the integration window to $1.2\,\mu$s. The resulting $\mu_1$ for FC only is $\mu_1 = 0.92 \pm 0.03$. With precise control of the peak width $\gamma$ and separation $\Delta$ (and thus the AFC finesse), we can keep the efficiency of the memory (and thus the total efficiency) quite constant for storage times up to $\tau_\mathrm{AFC} = 5\,\mu$s. The efficiency of the echo at a storage time of $\tau_\mathrm{AFC} = 1.6\,\mu$s sees an increase to $27.4 \pm 0.4\,\%$ due to the bandwidth of the pulses being narrower \cite{Moiseev2012}. The total efficiency for this storage time is $1.9 \pm 0.2\,\%$. At $\tau_\mathrm{AFC} = 5\,\mu$s where the echo efficiency is $\eta_{\mathrm{AFC}} = 19.1 \pm 0.2\,\%$, the $\mu_1$ is $8.6 \pm 0.5\times10^{-3}$ and the overall improvement seen is reduced to about 107 (compared to 270 for the previous case). This reduction is partly due to the increase in integration window (and hence noise) and because the pump gating was not as effective for this particular data set.

\section{Discussion}

Our conversion and storage device currently outperforms previous realizations of absorptive telecom memories at the single-photon-level \cite{Lauritzen2010, Lauritzen2011}. However the storage time and more so the efficiency demonstrated here needs significant improvement to be useful within quantum information processing technology. Despite this low efficiency operation, we have demonstrated that narrowband quantum memories can efficiently and effectively filter narrowband signals from broadband noise. Since only a fraction of the noise is stored in the memory, a drastic decrease in $\mu_1$ is achieved. Such filtering effects have been witnessed previously with AFC quantum memories in different contexts \cite{McAuslan2012, Rielander2014}. Rather than a telecom memory, our device can be viewed as a noiseless converter of telecom photons to the visible.

We now discuss strategies to improve the efficiency. There are 3 contributions to the total efficiency $\eta_\mathrm{tot}$: the device efficiency $\eta_\mathrm{dev}$ of the converter, the propagation loss $\eta_\mathrm{trans}$ from converter to memory, and the AFC memory efficiency $\eta_\mathrm{AFC}$. Below we discuss ways to improve these.

The device efficiency has 3 contributions, the internal efficiency, the signal coupling to the waveguide, and the transmission through filtering stages. Regarding the internal efficiency, we operated at a pump power much below $P_{\mathrm{max}}$  for the storage experiments due to limited available power. Therefore the total efficiency can be increased by simply increasing the pump power. Of course, an increased pump power necessarily results in an increased noise, however our memory can tolerate noise due to its effect of narrowband filtering. Note that the maximum internal efficiency is already quite high at $75\,\%$.

The coupling of the signal to the waveguide is only $55\,\%$ which could be increased with better mode-matching or fiber-pigtailing. Alternatively, a cavity-enhanced scheme with bulk crystal can be used  \cite{Albota2004, Vollmer2014}, where high internal efficiencies have also been observed.

The transmission of the filtering stage of the FC is $53\,\%$ most of which is due to the diffraction grating ($71\,\%$) and fiber coupling ($75\,\%$). This could be improved by using a volume-Bragg grating.

A major source of loss between the FC and the AFC is the depolarization of the light in the telecom fiber (slightly multimode at 606$\,$nm). We use 2 quarter-wave plates surrounding a half-wave plate to control the polarization giving around $80\,\%$ transmission through a polarizing beam splitter placed before the memory.

For the AFC efficiency, there are realizations that operated with higher efficiencies using $\mathrm{Pr}^{3+}$:$\mathrm{Y}_2\mathrm{Si}\mathrm{O}_5$. For example, using a cavity enhanced scheme as in \cite{Sabooni2013} an efficiency of $56\,\%$ was realized. Using an alternative memory scheme, $69\,\%$ efficiency was possible \cite{Hedges2010}.

We demonstrated that our device can store pulses of $560\,$ns duration which is compatible with the input pulses used for the full AFC memory demonstrated in \cite{Afzelius2010, Gundogan2013}. Therefore, in principle, a spin-wave AFC memory for up-converted telecom photons is possible. Using strong control pulses to transfer the  optical excitation to a spin-wave, the storage time can be increased to that of the hyperfine inhomogeneity (about $25\,\mu\mathrm{s}$ for $\mathrm{Pr}^{3+}$:$\mathrm{Y}_2\mathrm{Si}\mathrm{O}_5$  \cite{Afzelius2010, Gundogan2013}) and recalled on-demand. Further increasing of the storage time can be obtained via application of magnetic fields and dynamical decoupling techniques \cite{Fraval2005,Heinze2013}. We note that a spin-wave AFC memory is not possible in previous realizations of erbium-based telecom memories due to the non-compatible level structure.

One downside of our device is that the retrieved photon is still at $606\,$nm and would require yet another stage of frequency conversion to get back to the telecom wavelength, which would reduce the overall efficiency and increase greatly the possible $\mu_1$. However, for some quantum repeater architectures \cite{Sangouard2011}, it would not be necessary to convert back the stored photon.  Entanglement swapping operations that extend entanglement between neighbouring links in a quantum repeater can be done locally \cite{Sangouard2007a}, eliminating the need to convert back to telecom.

Previous experiments have shown that the process of non-linear frequency conversion \cite{Tanzilli2005, FernandezGonzalvo2013, Vollmer2014} and the AFC scheme \cite{Riedmatten2008, Clausen2011, Saglamyurek2011} preserve quantum superpositions. Our device could therefore be used to convert and store time-bin qubits. The high signal to noise ratio obtained in our experiment should allow for the overall fidelity to be in the quantum regime. However, our device is not directly compatible with polarization qubits as only one polarization (vertical) is guided. The memory alone has been shown to store arbitrary polarizations \cite{Gundogan2012}, so a device than can convert  arbitrary polarizations is needed to perform conversion and storage of polarization qubits. This could be realized with 2 waveguide set-ups, one converting the vertical component and one to convert the horizontal \cite{Vandevender2004}.

\section{Conclusion}

To conclude, we have demonstrated the storage of up-converted single-photon-level telecom light in a $\mathrm{Pr}^{3+}$:$\mathrm{Y}_2\mathrm{Si}\mathrm{O}_5$ crystal. We obtain an overall efficiency of $\eta_{\mathrm{tot}} = 1.9\pm0.2\,\%$ for a storage time of $1.6\,\mu$s. Storage times of up to $10\,\mu$s were also realized. This is the first demonstration of single-photon-level light storage interfaced with frequency up-conversion. Due to the fact that the echo is emitted at a predetermined time, and that the memory is narrowband relative to the broadband noisy conversion process, the signal to noise ratio of the retrieved light is 2 orders of magnitude superior to the input. With this in mind, we view our memory also as a noiseless converter of telecom light to the visible. Finally, our telecom conversion and storage device is in principle compatible with spin-wave storage protocols, a feature absent in previous realizations.

\section*{Acknowledgements}

We acknowledge financial support by the European projects CHIST-ERA QScale and FP7-CIPRIS (MC ITN-287252), by the ERC Starting grant QuLIMA, by the Spanish Ministry of Economy and Competitiveness (MINECO) and by the ``Fondo Europeo de Desarrollo Regional'' (FEDER) through grant FIS2012-37569.

\section*{References}
\bibliographystyle{ieeetr}

\end{document}